\documentstyle[ApJ,times,PSfig]{article}

\begin{document}

\def\simg{\mathrel{%
      \rlap{\raise 0.511ex \hbox{$>$}}{\lower 0.511ex \hbox{$\sim$}}}}
\def\siml{\mathrel{%
      \rlap{\raise 0.511ex \hbox{$<$}}{\lower 0.511ex \hbox{$\sim$}}}}
\def\Mesz{M\'esz\'aros~}
\def\ie{i.e$.$~} \def\eg{e.g$.$~} \def\etal{et al$.$~}
\def\eq{eq$.$~} \def\eqs{eqs$.$~} \def\deg{^{\rm o}} \def\dd{{\rm d}}
\def\beq{\begin{equation}} \def\eeq{\end{equation}}
\def\epsel{\varepsilon} \def\epselthird{\varepsilon_{-.5}}
\def\epsdy{\theta_{-1}} \def\xione{\xi_{-1}} \def\xifour{\xi_{-4}}

\title{Gamma-Ray Bursts from Up-Scattered Self-Absorbed Synchrotron Emission}

\author{A. Panaitescu}
\affil{Dept. of Astrophysical Sciences, Princeton University, Princeton, NJ 08544}
\and
\author{P. \Mesz}
\affil{ Institute for Advanced Study, Olden Lane, Princeton, NJ 08540,\\
       Institute of Astronomy, Madingley Road, Cambridge CB3 0HA, U.K.,\\
       Dept. of Astronomy \& Astrophysics, Pennsylvania State University, 
       University Park, PA 16802}

\begin{abstract}

We calculate the synchrotron self-Compton emission from internal shocks occurring 
in relativistic winds as a source of gamma-ray bursts, with allowance for self-absorption. 
For plausible model parameters most pulses within a Gamma-Ray Burst (GRB) are 
optically thick to synchrotron self-absorption at the frequency at which most 
electrons radiate. Up-scattering of photon number spectra harder than $\nu^0$ 
(such as the self-absorbed emission) yields inverse Compton photon number spectra 
that are flat, therefore our model has the potential of explaining the low-energy 
indices harder than $\nu^{-2/3}$ (the optically thin synchrotron limit) that 
have been observed in some bursts. The optical counterparts of the model bursts 
are sufficiently bright to be detected by such experiments as LOTIS, unless 
the magnetic field is well below equipartition.

\end{abstract}


\section{Introduction}

 In the framework of internal shocks occurring in highly relativistic winds, the 
up-scattering of synchrotron photons was taken into consideration in the calculation 
GRB spectra by Papathanassiou \& \Mesz (1996) and Pilla \& Loeb (1998). 
Both groups found a reasonable qualitative agreement between modeled and 
observed spectra. 
As shown in \S\ref{syic}, for electron injection fractions of order unity and  
reasonable dissipation efficiencies, synchrotron emission is liable to peak below 
the spectral peak observed in GRBs at hundreds of keVs (Band \etal 1993), even for 
equipartition magnetic fields. For this reason we consider a GRB model where the 
synchrotron peak is in/around the optical range, and the $\gamma$-photons arise from 
inverse Compton (IC) scatterings (\Mesz \& Rees 1994). In this case synchrotron 
self-absorption (SSA) can be important at the synchrotron peak, yielding up-scattered 
spectra that are harder below 100 keV than an optically thin synchrotron emission 
peaking at hard $X$-rays\footnote{If a photospheric component is present, then thermal 
and pair-Comptonized emission could also produce hard spectra (\Mesz \& Rees 2000)}.

 In this work we obtain numerically the synchrotron self-Compton emission from internal 
shocks through simulations of the wind dynamical evolution and the emission released in 
each collision. We also investigate the brightness of the optical flashes radiated by 
internal shocks (\Mesz \& Rees 1997).  
IC scattering of synchrotron seed photons has been previously considered as a mechanism
of emission of high energy photons in GRB afterglows (\eg \Mesz \& Rees 1994, Sari, 
Narayan, \& Piran 1996, Waxman 1997, Chiang \& Dermer 1999, Panaitescu \& Kumar 2000) 
and in blazars (\eg Boettcher, Mause, \& Schlickeiser 1997, Mastichiadis \& Kirk 1997, 
Urry 1998, Chiaberge \& Ghisellini 1999).

\section{Description of the Model}

\subsection{Wind Dynamics}
\label{wind}

The relativistic wind is approximated as a sequence of discrete shells of uniform
density, ejected by the GRB source with various initial Lorentz factors and masses.
We assume that the time it takes the GRB engine to become unstable and eject a
shell is proportional to the energy of that shell. This implies a constant wind 
luminosity $L$ on average throughout the entire wind ejection duration $t_w$. 
We consider that the shell Lorentz factors $\Gamma$ have a log-normal distribution
(Beloborodov 2000), $\Gamma-1 = (\Gamma_0-1) \exp(Ax)$ with $x$ Gaussian distributed,
and $A$ of order unity.

The radii at which the collisions between pairs of neighboring shells take place are 
calculated from the ejection kinematics, taking into account the progressive merging, 
the adiabatic losses between collisions, and the radiative losses. The Lorentz factor 
and internal energy of the shocked fluid are calculated from energy and momentum 
conservation in each shell collision. The shock jump conditions (Blandford \& McKee 1976)
determine the speeds of each shock, which give:
$i)$ the shocked fluid energy density $u'$ (used for calculating the magnetic field $B$ 
    and minimum electron Lorentz factor $\gamma_i$, see \S\ref{syic});
$ii)$ the shell shock-crossing time $t'_i$ (which is the duration of the injection of 
      electrons, determining the cooling electron Lorentz factor $\gamma_c$, see
      \S\ref{eldist}); and
$iii)$ the post-shock shell thickness $\Delta'$ (used for calculating the adiabatic 
       cooling timescale and the radiative efficiency). 
Between collisions the co-moving frame thickness increases at the sound speed.

\subsection{Emission of Radiation} 
\label{syic}

The synchrotron self-Compton emission from each shocked shell is calculated from the 
Lorentz factor, magnetic field, number of electrons (given by the shell mass) and the 
electron distribution in each shocked shell. The turbulent magnetic field strength $B$ 
and the typical electron Lorentz factor $\gamma_i$ resulting from shock-acceleration are 
parameterized through their fractional energy $\xi$ and $\epsel$ relative to the 
internal energy:
\beq
 B^2 = 8\pi\,\xi\,u' \quad,\quad 
 \gamma_i = \frac{p-2}{p-1} \frac{\epsel\,u'}{n'_e\,m_e c^2} \;,
\label{Bgi}
\eeq
where $u'$ and $n'_e$ are the internal energy and particle density\footnote{
   Primed quantities are in the co-moving frame. The values used in the numerical 
   calculations are those when the shock has swept up the entire shell.},
respectively, and $p$ is the index of the power-law distribution of injected electrons:
${\cal N}_i(\gamma) \propto \gamma^{-p}$ for $\gamma \geq \gamma_i$. Denoting by 
$\theta$ the laboratory frame internal-to-kinetic energy ratio in the shocked fluid 
(\ie the dissipation efficiency), one can write $u' \simeq \theta\, n'_e m_p c^2$, 
hence
\beq 
 \gamma_i = ({m_p}/{2\,m_e})\; \epsel\, \theta = 30\; \epselthird\, \epsdy \;,
\label{gi}
\eeq
with $A_n \equiv 10^{-n} A$ and using $p=3$ for simplicity.

The density $n'_e$ can be calculated from the shell mass $m$, thickness $\Delta'$, and 
radius $r$, using $m = 4\pi r^2 m_p  n'_e \Delta'$. For analytical calculations one can 
approximate the typical mass $m$ of a shell radiating a pulse which arrives at observer 
time $T$ as a fraction $T/T_b$ of the total wind mass $M$, where $T_b$ is the burst 
duration, which is approximately equal to the duration $t_w$ of the wind ejection. 
One obtains $m = LT/\Gamma c^2$, where $L \equiv E/t_w$ is the wind luminosity, 
$E$ and $\Gamma$ being the wind total energy average Lorentz factor of the wind. 
The collisional radius $r$ can be derived from the spread in the photon arrival 
time $\delta = r/\Gamma^2 c$ due to the spherical shape of the emitting surface, which 
is roughly the pulse duration\footnote{
   The true pulse duration is larger, due to contributions from the shell shock-crossing 
   and the cooling time (radiative and adiabatic) of the electrons radiating at the 
   observing frequency. These are taken into account numerically and influence the
   light-curves and instantaneous spectra, but not the burst-integrated spectra.}.
The shell thickness can be approximated\footnote{ 
   This is a good approximation for a freely expanding shell; shock-compression and 
   shell merging yield smaller shell thicknesses.}
by $\Delta' = r/\Gamma$. Therefore we obtain $n'_e  = LT/(4\pi m_p c^5 \Gamma^6 \delta^3)$. 
Then equation (\ref{Bgi}) yields
\beq
 B = \frac{ \left( 2\,\xi \theta LT \right)^{1/2} }{ \Gamma^3 (c \delta)^{3/2} } 
     = 3\times 10^4 \; \xione^{1/2}\,\epsdy^{1/2}\,L_{53}^{1/2}\,\Gamma_2^{-3}
       \,\delta_0^{-3/2} T_1^{1/2} \; {\rm (G)} \;.  
\label{Bmag}
\eeq
If the emission at $\nu_i$ is not self-absorbed, then the peak frequency $\nu_{sy}$ of
the synchrotron power $\nu F_\nu^{(sy)}$ is at the characteristic synchrotron frequency
$\nu_i = (eB \gamma_i^2 \Gamma)/(4\, m_e c)$ of the $\gamma_i$-electrons, and the
peak of the IC power $\nu F_\nu^{(ic)}$ is at $\nu_{ic} \simeq  \gamma_i^2 \nu_i$.
From equations (\ref{gi}) and (\ref{Bmag}), one obtains that
\beq
 \nu_{sy} = 10^{16}\, \epselthird^2 \xione^{1/2} \epsdy^{5/2} L_{53}^{1/2}
                   \Gamma_2^{-2}\, \delta_0^{-3/2} T_1^{1/2} \; {\rm (Hz)} \;,
\label{nui}
\eeq
\beq
 \nu_{ic} \simeq 10^{19}\, \epselthird^4 \xione^{1/2} \epsdy^{9/2}
            L_{53}^{1/2} \Gamma_2^{-2}\, \delta_0^{-3/2} T_1^{1/2} \; {\rm (Hz)} \;.
\label{nuii}
\eeq

 Equations (\ref{nui}) and (\ref{nuii}) show that the peak frequencies of the optically
thin synchrotron and of the IC emissions depend strongly on parameters which may vary 
substantially from pulse to pulse. Equation (\ref{nui}) also shows that for plausible
model parameters the synchrotron peak $h\nu_{sy}$ cannot be as high as 100 keV for
pulses longer than about $\delta = 0.1\,{\rm s}$.

 Due to the collisions among shells, the amplitude of the Lorentz factor fluctuations 
in the wind is reduced, diminishing the dissipation efficiency $\theta$. Equation 
(\ref{nuii}) shows that the peak of the IC emission depends strongly on $\theta$ and 
thus should have a general trend of decreasing in time. This could be the reason for the 
overall spectral softening seen in GRBs (\eg Ford \etal 1995).

\subsection{Electron Distribution}
\label{eldist}

The electron distribution ${\cal N} (\gamma)$ at the end of shell energization is 
determined by the ${\cal N}_i (\gamma)$ initially injected at shock, by the electron 
radiative cooling (through synchrotron and IC radiation) and by the absorption of the 
synchrotron photons. The radiative cooling timescale of electrons with Lorentz factor 
$\gamma$ is $t'_r(\gamma) \sim (\gamma m_e c^2)/(P'_{sy}+P'_{ic})$, where $P'_{ic} =
Y P'_{sy}$ is the electron IC radiating power and $P'_{sy} = (1/6\pi) \sigma_{Th} c
\gamma^2 B^2$ is the synchrotron power, thus
\beq
 t'_r(\gamma) = \frac{C_1}{(Y+1)\gamma} t'_i\;, 
           \quad C_1 \equiv \frac{6\pi m_e c}{\sigma_{Th} B^2 t'_i} \;,
\label{trad}
\eeq
$t'_i$ being the time elapsed since the beginning of injection of relativistic electrons, 
\ie the shell shock-crossing time, which is calculated numerically from the shock dynamics.

From equation (\ref{trad}) the cooling electron Lorentz factor $\gamma_c$ defined
by $t'_r(\gamma_c) = t'_i$ is $\gamma_c = C_1/(Y+1)$. For $\gamma_c < \gamma_i$ the
$\gamma_i$-electrons cool faster than their injection timescale. 
In this {\sl fast electron cooling} regime the resulting electron distribution has 
a low-energy tail ${\cal N} (\gamma) \propto \gamma^{-2}$ for $\gamma_c < \gamma <
\gamma_i$, while ${\cal N} (\gamma) \propto \gamma^{-(p+1)}$ above $\gamma_i$.
In the {\sl slow electron cooling} regime, characterized by $\gamma_i < \gamma_c$, 
the electron distribution is the injected one, ${\cal N} (\gamma) \propto \gamma^{-p}$, 
for $\gamma_i < \gamma < \gamma_c$, and  ${\cal N} (\gamma) \propto \gamma^{-(p+1)}$ 
above $\gamma_c$, due to cooling. As physical parameters vary among shocked shells, 
the GRB pulses can be in different electron cooling regimes.

\subsection{Compton Parameter and Self-Absorption Break}
\label{Yab}

 We set the electron distribution as described in \S\ref{eldist}. The Compton parameter 
is given by 
\beq
 Y = \tau_e \left[ 
      \int_0^{\gamma_a} {\cal N}(\gamma) \frac{\gamma^2}{\tau_a(\gamma)} {\rm d}\gamma + 
      \int_{\gamma_a}^\infty {\cal N}(\gamma) \gamma^2 {\rm d}\gamma 
            \right]\;,
\label{Ypar}
\eeq
where $\tau_e$ is the optical depth to electron scattering and $\tau_a(\gamma)$ is the 
optical thickness to SSA at the synchrotron characteristic frequency for $\gamma$-electrons, 
and $\gamma_a$ is defined by $\tau_a(\gamma_a) =1$. The first integral in the right-hand 
side of equation (\ref{Ypar}) takes into account that only those photons emitted within a 
region (around a given electron) of optical depth unity are up-scattered before being 
absorbed.  Table 1 lists the values of the Y parameter for all possible orderings of 
$\gamma_a$, $\gamma_c$, and $\gamma_i$, assuming that these breaks are sufficiently apart 
from each other, and ignoring multiplicative factors of order unity.

For a power-law distribution of electrons it can be shown that the optical thickness
$\tau_a$ to SSA at the characteristic synchrotron frequency for electrons at the top of 
the distribution, \ie at $\gamma_p = \min(\gamma_c,\gamma_i)$, is 
\beq 
 \tau_p \simeq \frac{C_2}{\gamma_p^5} \;, \quad
      C_2 \equiv \frac{5\, e\tau_e}{\sigma_{Th} B} \;.
\label{taup}
\eeq
For slow electron cooling ($\gamma_p = \gamma_i$) one obtains
\beq
  \tau_p = 3 \times 10^4 \; \epselthird^{-5} \xione^{-1/2} \epsdy^{-11/2}
                 L_{53}^{1/2} \Gamma_2^{-2} \delta_0^{-1/2} T_1^{1/2} \;.
\label{taui}
\eeq
If electrons are cooling fast ($\gamma_p = \gamma_c$), then $\tau_p$ is even larger.
Equation (\ref{taui}) shows that is quite likely that the synchrotron emission from 
the typical, $\gamma_p$-electron is self-absorbed, thus the synchrotron spectrum 
$F_\nu$ peaks at the synchrotron frequency $\nu_a$ corresponding to $\gamma_a$.

The shell optical thickness to SSA at any frequency $\nu$ (or corresponding $\gamma$) 
can be calculated from $\tau_p$ given in equation (\ref{taup}) by using 
$\tau_a (\nu) \propto \nu^{-5/3} \propto \gamma^{-10/3}$ for $\gamma < \gamma_p$ and 
$\tau_a (\nu) \propto \nu^{-(q+4)/2} \propto \gamma^{-(q+4)}$ for $\gamma_p < \gamma$, 
where $q$ is the index of the electron distribution around $\gamma$, \ie 
$q = 2$ for $\gamma_c < \gamma < \gamma_i$, $q = p$ for $\gamma_i < \gamma < \gamma_c$, 
and $q = p+1$ for $\max(\gamma_i,\gamma_c < \gamma$.

To take into account that for $\gamma_c \ll \gamma_a$ the synchrotron cooling is 
basically suppressed by self-absorption, we calculate $\gamma_c$ from $\gamma_c = 
C_1/[Y+\chi(\gamma_c/\gamma_a)]$, where $\chi$ is a function satisfying $\lim_{x\to0} 
\chi(x) = 0$, $\chi(1)=1-e^{-1}$, and $\lim_{x\to\infty} \chi(x) = 1$.  
Note that the Y parameter and the self-absorption $\gamma_a$ depend on the electron 
distribution, which is at its turn determined by the electron cooling through 
synchrotron (possibly reduced by self-absorption) and IC. Thus the equations for 
$\gamma_a$, $Y$ (see Table 1), and $\gamma_c$ are coupled. For given parameters
$(C_1,C_2,\tau_e,\gamma_i,p)$ these equations are solved numerically for each of the 
cases listed in Table 1 and only the self-consistent solution (\ie the one that 
satisfies the assumed ordering of $\gamma_a$, $\gamma_c$, and $\gamma_i$) is retained 
for the calculation of the synchrotron and IC emissions.

\subsection{Synchrotron and Inverse Compton Spectra}
\label{absic}

The synchrotron spectrum $F_\nu^{(sy)}$ is approximated as a sequence of four 
power-laws with breaks at $\nu_a$, $\nu_i$, and $\nu_c$. The slope of the spectrum 
between break depends on the ordering of these breaks: 
$F_\nu^{(sy)} \propto \nu^{5/2}$ for $\min (\nu_c,\nu_i) < \nu < \nu_a$, 
$F_\nu^{(sy)} \propto \nu^2$ for $\nu < \min (\nu_a,\nu_c,\nu_i)$, 
$F_\nu^{(sy)} \propto \nu^{1/3}$ for $\nu_a < \nu < \min (\nu_c,\nu_i)$, 
$F_\nu^{(sy)} \propto \nu^{-1/2}$ for $\max (\nu_a,\nu_c) < \nu < \nu_i$, 
$F_\nu^{(sy)} \propto \nu^{-(p-1)/2}$ for $\max (\nu_a,\nu_i) < \nu < \nu_c$, and 
$F_\nu^{(sy)} \propto \nu^{-p/2}$ for $\max (\nu_a,\nu_c,\nu_i) < \nu$.

The up-scattered spectrum has breaks and slopes that are determined by those of the 
synchrotron spectrum $F_{\nu}^{(sy)}$ and of the electron distribution ${\cal N}
(\gamma)$, thus the IC spectrum is more complex than that of the synchrotron emission. 
Figure 1 shows the logarithmic derivatives (\ie slopes) of the IC emission for all
possible orderings of the synchrotron breaks, obtained by integrating numerically over 
$F_{\nu}^{(sy)}$ and ${\cal N}(\gamma)$ the up-scattered emission per electron given 
in equation (2.48) of Blumenthal \& Gould (1970). Note that the IC spectrum resulting 
from up-scatterings of photons below $\nu_a$, where $F_\nu^{(sy)} \propto \nu^{5/2}$ or
$F_\nu^{(sy)} \propto \nu^2$, by a typical electron with $\gamma = \min (\gamma_c,
\gamma_i)$ cannot be harder than $F_\nu^{(ic)} \propto \nu$.

Numerically, we calculate the IC spectrum by approximating it as a sequence of 
power-laws, with breaks at the frequencies identified in Figure 1, and we ignore 
higher order IC scatterings.  For the model parameters we shall consider, scatterings 
of third order or higher are suppressed as they occur in extreme Klein-Nishina regime. 
However this is not always true for second IC scatterings. Therefore our calculations 
ignore a very high energy (above 1 GeV) component, and may overestimate the brightness 
of the synchrotron and first IC components (but not their ratio) as some electron energy 
would be ``drained" through a second up-scattering.

\section{Simulated Bursts}
\label{spek}

The burst spectrum is calculated by summing the synchrotron and IC spectra from all 
collisions.  Figure 2 shows the time-integrated spectrum $\Phi_\nu = \int F_\nu(T) {\rm d}T$ 
for a wind consisting of 13 shells, for plausible model parameters $(L,t_w;\epsel, \xi)$, 
and for a log-normal distribution of $\Gamma$. The burst has two major structures whose
spectra are shown separately. The first one ($0-5\,{\rm s}$) consists of two pulses whose 
IC emissions peak at 10 keV and 2 MeV and has a rather soft spectrum in the BATSE range. 
The second structure ($15-20\,{\rm s}$) is a single pulse, self-absorbed at the synchrotron
peak, with the IC emission peaking at 200 keV and having a hard spectrum below 40 keV.
The other six collisions occurring in the wind considered here are very dim in the 
10 keV--1 MeV range either because their radiative efficiency is low or because their
synchrotron and IC emissions peak far from this photon energy range.

Other sets of parameters may also lead to hard low energy slopes resulting from up-scattering
of self-absorbed synchrotron emission. However, since the physics of the model does not
confine the peak frequencies IC emission of the bright pulses within a decade, the spectra 
resulting from the addition of many pulses do not exhibit in general a steep slope below 
and around 10 keV. Therefore our model for the GRB emission can explain the hard low energy 
spectra observed by Preece \etal (1998) only for bursts with a modest number of pulses.

Figure 3 shows the dependence of the $R$-band magnitude of the burst optical counterpart 
on the most important model parameters. The wind parameters used in these calculations
produce GRBs of average intensity and with spectra compatible with the observations.
Notice that most of the optical counterparts are above the reported sensitivity of the 
up-graded LOTIS experiment (Williams \etal 2000). The exception is the case of a wind with 
a very low magnetic field parameter $\xi$. 
 
The non-detection of such optical counterparts can be used to infer lower limits on the IC 
parameter, which sets the relative intensity of the high ($\gamma$-ray) and low (optical)
frequency emission. It is straightforward to show that a ``typical" GRB lasting $10\,{\rm s}$, 
with a $\nu F_\nu$ peak frequency around 100 keV and fluence of $5\times10^{-6}\,{\rm erg\, 
cm^{-2}}$, has an $R$-band 
magnitude
\beq
 R = 8.8 + 2.5\, (1-\beta) \log X + 2.5\, \log Y_2 \;,
\label{Rmag}
\eeq  
where $X$ is the ratio of the peak frequency of the synchrotron power $\nu F_\nu^{(sy)}$ 
to that of observations ($\nu_R \sim 4.7 \times 10^{14}$ Hz), and $\beta$ is the slope of 
$F_\nu^{(sy)}$, \ie $F_\nu \propto \nu^{-\beta}$. Equation (\ref{Rmag}) shows that, if the
synchrotron peak frequency is in or not too far from the optical domain, as is the case
for the winds whose optical light-curves are shown in Figure 3, then the average 
$R$-magnitude of the optical flash is dimmer than $R_{LOTIS} \sim 13$ (Williams \etal 
1999) if $Y \simg 5000$.

For fast electron cooling and in the $\nu_a < \nu_i$ case, \ie for the first two 
cases\footnote{ These cases are encountered in a fraction of the simulated pulses 
       for the parameters given in Figure 3, nevertheless the increase of $Y$ with 
       decreasing $\xi$ is a general feature exhibited by the simulated bursts.} 
given in Table 1, the electron distribution of \S\ref{eldist} yields $Y \simeq \gamma_c 
\gamma_i \tau_e$ which, in the $\xi \ll \epsel$ limit, leads to
\beq
 Y = (C_1 \gamma_i \tau_e)^{1/2} = \left[ \frac{(3/4)(\Delta'/t'_i) n'_e m_e c^2 \gamma_i}
       {u'_B} \right]^{1/2} = \left( \frac{3\,\epsel}{4\,\xi} \right)^{1/2} \;, 
\label{Yr}
\eeq
where $u'_B=B^2/8\pi$ is the magnetic energy density, and $\Delta' \sim c t'_i$ was used.
The above condition for LOTIS non-detection ($Y \simg 5000$) implies $\xi \siml 10^{-8}\, 
\epselthird$.  Therefore the optical flash of a ``typical" burst would be dimmer than about 
$R=13$ if the magnetic field is several orders of magnitude below equipartition\footnote{ 
 We note that if only a small fraction of the injected electrons were to acquire the 
 fractional energy $\epsel$ and if the magnetic field is close to equipartition 
 ($\xi \siml 1$), such that the synchrotron emission would peak around 100 keV, 
 the optical flashes would also be dimmer than the LOTIS limit.}.

\section{Conclusions}

The analytical treatment presented in \S\ref{syic} and the numerical spectrum displayed 
in Figure 2 show that IC up-scattering of synchrotron emission from internal shocks 
in relativistic unstable winds can produce GRBs with break energies and low- and high-energy 
indices that are typical of real GRBs (Preece \etal 2000). The peak frequency of the IC 
emission is strongly dependent on some of the model parameters (see \eq [\ref{nuii}] for 
the first two cases listed in Table 1). For a range of plausible parameters this peak 
falls within the BATSE observing window. 

The ``harder than synchrotron" low-energy spectra that have been reported in a significant 
fraction of bursts (Preece \etal 1998) can be explained by the Compton up-scattering of 
synchrotron spectra that are self-absorbed. Even though the synchrotron spectrum can be as
hard as $F_\nu^{(sy)} \propto \nu^{5/2}$ below the absorption break, the up-scattered 
spectrum cannot be harder than $F_\nu^{(ic)} \propto \nu^1$, therefore the self-absorbed 
self-Compton model has a limiting value $\alpha_{ic} \leq 0$ (the low energy index of the 
photon spectrum). This is larger by $\delta \alpha = 2/3$ than the limiting value for an 
optically thin synchrotron spectrum, leaving only a couple of bursts in Figure 2 of Preece 
\etal (1998) with $\alpha$'s exceeding $\alpha_{ic}$ by more than $1\sigma$.
We emphasize that self-absorption of the synchrotron emission is not a guarantee that the
resulting model spectrum is as a hard as $\alpha_{ic} = 0$, as the addition of the IC 
emission of many pulses with various IC peak frequencies may result in a flatter burst 
spectrum, and that, in general, hard low energy spectra can be obtained only for bursts
with a few pulses.

The prompt optical to gamma-ray emission ratio from internal shocks depends on the 
fractional energy in the magnetic field -- a poorly known parameter. This is independent 
and additional to a possible optical flash from a reverse component of a subsequent 
external shock (\Mesz \& Rees 1997, Sari \& Piran 1999). A possible explanation for 
the non-detection of optical counterparts down to $R \sim 13$ by LOTIS is that the 
magnetic field in internal shocks  is several orders of magnitude weaker than the 
equipartition value, corresponding to an average Compton parameter around or larger 
than 1000.

\acknowledgements{We are grateful to B. Paczy\'nski and M.J. Rees for comments. 
AP acknowledges support from a Lyman Spitzer Jr. fellowship, and PM from NASA 
NAG-5 9192, the Guggenheim Foundation, the Institute for Advanced Study and 
the Sackler Foundation}

\newpage

\begin{table}
\begin{center}
 TABLE 1. \\ {\sc The Compton Parameter ($Y$), self-absorption Lorentz factor ($\gamma_a$) 
              and peak of \\ the IC $\nu F_\nu$ spectrum ($\nu_{ic}$) for various possible 
              orderings of $\gamma_a$, $\gamma_c$, and $\gamma_i$} \\ [4ex]
 \begin{tabular}{cccc} \hline \hline
 \rule[-4mm]{0mm}{10mm}   case  &  $Y/\tau_e$  &  $\gamma_a$  &  $\nu_{ic}$  \\ 
               \hline \hline
 \rule[-2mm]{0mm}{7mm}  $\gamma_a < \gamma_c < \gamma_i$  &  $\gamma_c \gamma_i$  &
                          $C_2^{3/10} \gamma_c^{-1/2}$ &  $\gamma_i^2 \nu_i$  \\
 \rule[-2mm]{0mm}{7mm}  $\gamma_c < \gamma_a < \gamma_i$  &  $\gamma_c \gamma_i$ &
                          $(C_2 \gamma_c)^{1/6}$ &  $\gamma_i^2 \nu_i$  \\
 \rule[-2mm]{0mm}{7mm} $\gamma_c<\gamma_i<\gamma_a$ & $\gamma_c \gamma_i^{p-1} \gamma_a^{2-p}$ &
                          $(C_2 \gamma_c \gamma_i^{p-1})^{1/(p+5)}$ &  $\gamma_i^2 \nu_a$  \\ 
                   \hline
 \rule[-2mm]{0mm}{9mm}  $\gamma_a < \gamma_i < \gamma_c$ & $\left\{ \begin{array}{ll}  
              \gamma_i^{p-1} \gamma_c^{3-p} & p < 3 \\ \gamma_i^2  & p > 3 \end{array} \right.$ &
                          $C_2^{3/10} \gamma_i^{-1/2}$ & $\left\{ \begin{array}{ll} 
               \gamma_c^2 \nu_c & p < 3 \\ \gamma_i^2 \nu_i & p > 3 \end{array} \right.$  \\
 \rule[-2mm]{0mm}{9mm}  $\gamma_i < \gamma_a < \gamma_c$ & $\left\{ \begin{array}{ll}  
   \gamma_i^{p-1} \gamma_c^{3-p} & p < 3 \\ \gamma_i^{p-1} \gamma_a^{3-p} & p > 3 \end{array} \right.$ &
                       $(C_2 \gamma_i^{p-1})^{1/(p+4)}$ & $\left\{ \begin{array}{ll} 
               \gamma_c^2 \nu_c & p < 3 \\ \gamma_i^2 \nu_a & p > 3 \end{array} \right.$  \\
 \rule[-2mm]{0mm}{9mm}  $\gamma_i < \gamma_c < \gamma_a$ & $\gamma_i^{p-1} \gamma_c \gamma_a^{2-p}$ &
                       $(C_2 \gamma_i^{p-1} \gamma_c)^{1/(p+5)}$ & $\left\{ \begin{array}{ll} 
                \gamma_c^2 \nu_a & p < 3 \\ \gamma_i^2 \nu_a & p > 3 \end{array} \right.$  \\ 
                   \hline \hline
 \end{tabular}
\end{center}
\end{table}

\begin{figure*}
\centerline{\psfig{figure=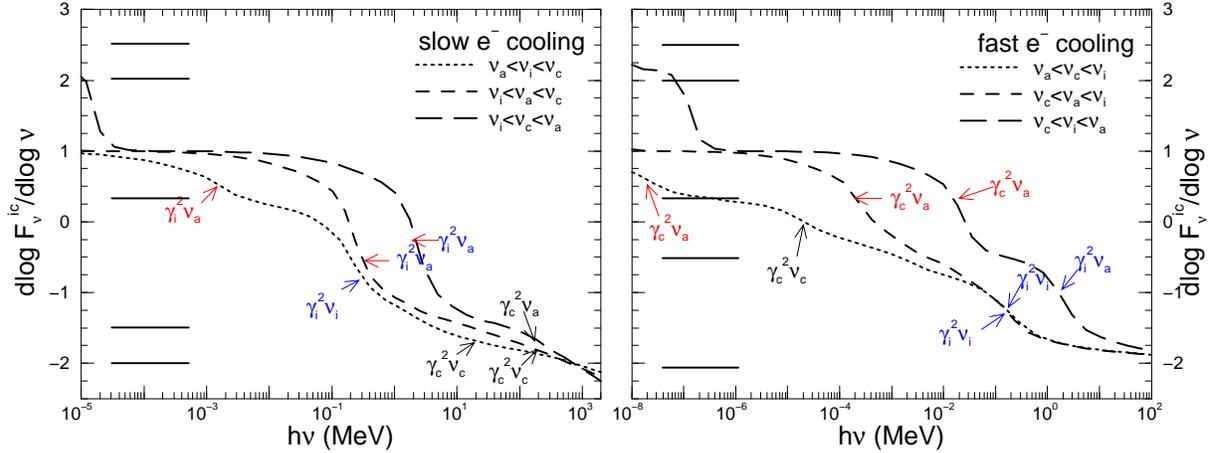}}
\figcaption{Slopes of IC spectra resulting from up-scattering of a broken power-law 
 (four segments) synchrotron spectrum by a broken power-law distribution of electrons, 
 for various possible combinations of synchrotron breaks: $\nu_a$ (absorption), $\nu_i$ 
 (injection), and $\nu_c$ (cooling). Horizontal lines indicate the slopes of the 
 synchrotron spectrum, which may be asymptotic values for the IC spectrum slope. 
 Note that the IC emission below the lowest energy break indicated for each spectrum 
 (red arrows), which results from up-scattering of self-absorbed synchrotron emission, 
 is $F_{\nu}^{(ic)} \propto \nu$. Blue arrows/characters indicate the IC break frequency 
 closest to the peak of the $\nu F_\nu^{(ic)}$ emission. Black arrows indicate less 
 conspicuous breaks or IC frequencies that separate two asymptotic values 
 of the IC spectral slope. A steep injected electron distribution with $p=4$ was used
 so that the IC breaks are easier to identify. IC spectra were calculated in the co-moving 
 frame and blue-shifted by a factor 100.}
\end{figure*}

\begin{figure*}
\hspace*{1cm}
\centerline{\psfig{figure=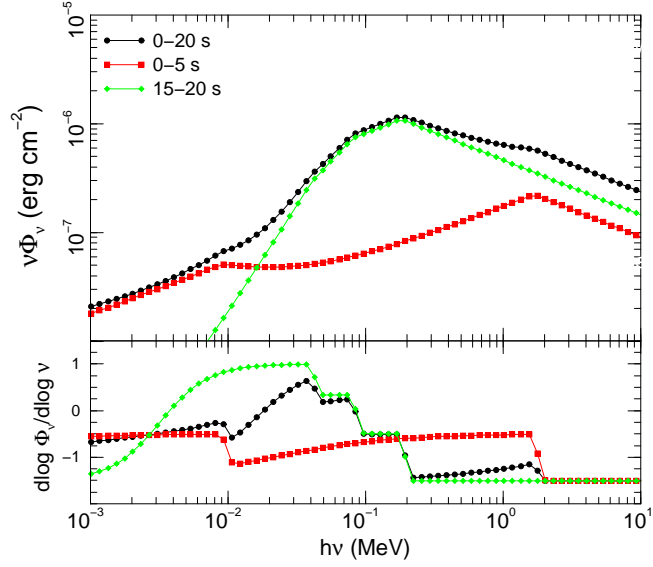}}
\figcaption{Time-integrated spectrum for a simulated burst produced by a wind consisting
 initially of 13 shells, whose ejection Lorentz factors has a log-normal distribution
 with $\Gamma_0=320$ and $A=1$ (see \S\ref{wind}). Other wind parameters are: $t_w = 10$ s, 
 $L = 3\times 10^{52}\; {\rm erg\, s^{-1}}$, $\epsel = 0.15$, $\xi = 10^{-5}$, $p=3$, redshift 
 $z=1$ (see \S\ref{syic}). The upper panel shows separately the spectra of the two structures 
 exhibited by the light-curve, identified by the time ranges when they are seen, as well as 
 the overall burst spectrum. Note the steepness of the $15-20\,{\rm s}$ spectrum below 
 40 keV peak (lower panel), due to up-scattering of self-absorbed synchrotron emission.
 The average dissipation efficiency per collision is $\theta=13\%$. Approximately 29\% of 
 the wind energy is eventually dissipated, out of which $\sim$ 4\% is stored in electrons. 
 About 50\% of the electron energy is lost adiabatically, therefore the bolometric radiative 
 efficiency of the burst is only 2.0\%. The efficiency at which the wind converts its kinetic 
 energy into 20 keV--1 MeV photons is 1.2\%, due to that only 60\% of the total emission falls 
 in this energy range.}
\end{figure*}  

\begin{figure}
\vspace*{-1cm} \hspace*{-1cm}
\centerline{\psfig{figure=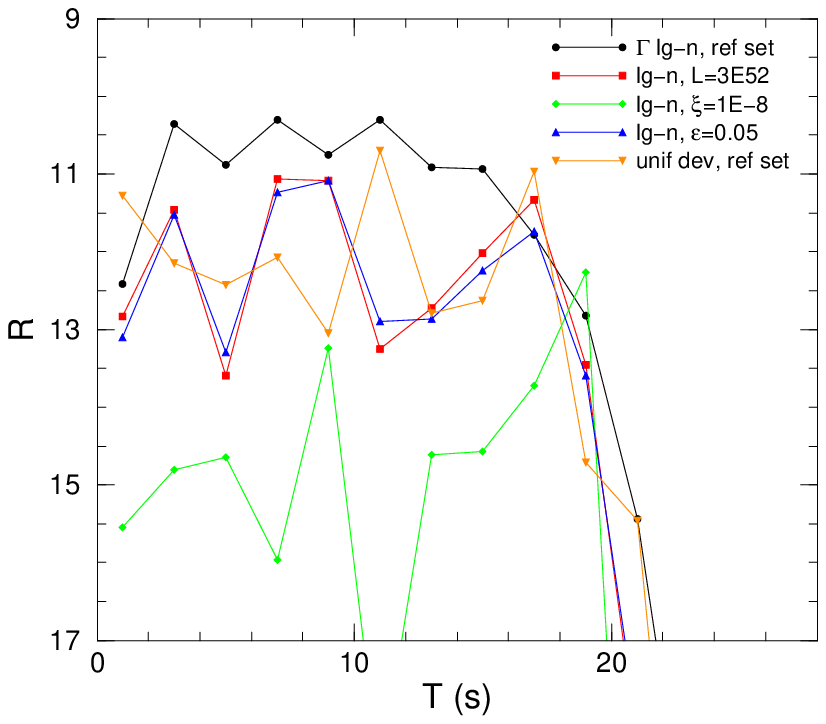}}
\figcaption{$R$-band magnitude of burst optical counterparts for few model parameters,
 bined in $2\,{\rm s}$ time intervals. The reference set is $t_w=10$ s, $L=10^{53}\; 
 {\rm erg\, s^{-1}}$, $\epsel = 0.15$, $\xi=10^{-5}$, $p=3$, redshift $z=1$ and 100 ejected 
 shells. $\Gamma_0=320$ and $A=1$ for $\Gamma$ log-normal distributed, while for the
 uniform deviate $\Gamma$ is randomly distributed between 50 and $10^3$. The bursts last for
 about $(1+z)t_w=20$ s, exhibit 20--35 peaks, have spectral features (low and high energy 
 spectral slope, break frequency) consistent to those found by Preece \etal 2000 in real 
 bursts, 20 keV--1 MeV fluences in the range $(2-8)\times 10^{-6}\;{\rm erg\, cm^{-2}}$, and 
 efficiencies of radiating the wind total energy in the BATSE range between 0.3\% and 1\%.}
\end{figure}


\begin{references}

\reference{} Band, D. \etal 1993, ApJ, 413, 281
\reference{} Beloborodov, A 2000, ApJ, 539, L25
\reference{} Blandford, R.D. \& McKee, C.F. 1976, Phys of Fluids, vol 19, no 8, 1130
\reference{} Blumenthal, G. \& Gould, R. 1970, Rev. Mod. Phys., 42, no 2, 237
\reference{} Boettcher, M., Mause, H., \& Schlickeiser, R. 1997, A\&A, 324, 395
\reference{} Chiaberge, M. \& Ghisellini, G. 1999, MNRAS, 306, 551
\reference{} Chiang, J. \& Dermer, C.D. 1999, ApJ, 512, 699
\reference{} Ford, L.A. \etal 1995, ApJ, 439, 307
\reference{} Mastichiadis, A. \& Kirk, J.G. 1997, A\&A, 320, 19
\reference{} \Mesz, P. \& Rees, M.J. 1994, 269, L41
\reference{} \Mesz, P. \& Rees, M.J. 1997, 476, 232
\reference{} \Mesz, P. \& Rees, M.J. 2000, ApJ, 530, 292
\reference{} Panaitescu, A. \& Kumar, P. 2000, ApJ, 543, in press (astro-ph/0003246)
\reference{} Papathanassiou, H. \& \Mesz, P. 1996, ApJ, 471, L91 
\reference{} Pilla, R. \& Loeb, A. 1998, 494, L167
\reference{} Preece, R., Briggs, M., Mallozzi, R., Pendleton, G., Paciesas, W., 
             \& Band, D. 1998, ApJ, 506, L23
\reference{} Preece, R., Briggs, M., Mallozzi, R., Pendleton, G., Paciesas, W., 
             \& Band, D. 2000, ApJS, 126, 19
\reference{} Sari, R., Narayan, R., \& Piran, T. 1996, ApJ, 473, 204
\reference{} Sari, R. \& Piran, T. 1999, ApJ, 517, L109
\reference{} Urry, C.M. 1998, in Adv. in Space Res., vol 21, vol 1-2, p.89
\reference{} Waxman, E. 1997, ApJ, 485, L5
\reference{} Williams, G. \etal 1999, ApJ, 519, L25
\reference{} Williams, G. \etal 2000, in Proc. of the $5^{th}$ Huntsville GRB Symposium \\
             (astro-ph/9912402) 

\end{references}
\end{document}